\documentclass[jkps,preprint,fleqn,showpacs,showkeys]{revtex4}
\usepackage{graphicx}
\usepackage{amssymb}
\usepackage{amsmath}
\usepackage{bm}
\begin{document}
\setcounter{page}{0}
\title[]{Development of the Magnetic Excitations of Charge-Stripe Ordered  La$_ {2-x}$Sr$_ x$NiO$_ 4$ on Doping Towards Checkerboard Charge Order.}
\author{P. G. \surname{Freeman}}
\email{paul.freeman@epfl.ch}
\affiliation{Laboratory for Quantum Magnetism, Ecole Polytechnique Federale de Lausanne (EPFL), Switzerland.}
\affiliation{Helmholtz-Zentrum Berlin n f\"{u}r Materialien und Energie GmbH, Hahn-Meitner-Platz 1, D-14109 Berlin, Germany}
\affiliation{Institut Laue-Langevin, BP 156, 38042 Grenoble Cedex 9, France.}

\author{S. R.  \surname{Giblin}}
\affiliation{ISIS Facility, Rutherford Appleton Laboratory, Chilton,
Didcot, Oxon, OX11 0QX, UK.}
\affiliation{Cardiff School of Physics and Astronomy, Cardiff University, Queens Buildings, The Parade, Cardiff, CF24 3AA, UK.}

\author{K. \surname{Hradil}}
\affiliation{Institut f\"{u}r Physikalische Chemie, Universit\"{a}t
G\"{o}ttingen, Tammanstrasse 6, 37077 G\"{o}ttingen, Germany}
\affiliation{Technische Universit�t Wien,
Karlsplatz 13, 1040 Vienna, Austria}

\author{R. A.  \surname{Mole}}
\affiliation{Bragg Institute, ANSTO, New Illawarra Road, Lucas Heights, NSW,  Australia}
\affiliation{FRM II, Lichtenbergstra\ss e, 1 85747 Garching, Germany}

\author{D. \surname{Prabhakaran}}
\affiliation{Department of Physics, University of Oxford, Oxford, OX1
3PU, UK}

\date[]{Received 30 April 2011}

\begin{abstract}
The magnetic excitation spectrums of charge stripe ordered La$_{2-x}$Sr$_{x}$NiO$_4$ $x = 0.45$ and $x = 0.4$ were studied by inelastic neutron scattering.
We found the magnetic excitation spectrum of  $x = 0.45$ from the ordered Ni$^{2+}$ S = 1 spins to match that of checkerboard charge ordered La$_{1.5}$Sr$_{0.5}$NiO$_4$. The distinctive asymmetry in the magnetic excitations above 40\,meV was observed for both doping levels,  but an additional ferromagnetic mode was observed in $x = 0.45$ and not in the  $x = 0.4$. We discuss the origin of crossover in the excitation spectrum between  $x = 0.45$ and $x = 0.4$ with respect to discommensurations in the charge stripe structure. 
\end{abstract}

\pacs{71.45.Lr, 75.30.Fv, 75.40.Gb}

\keywords{Stripe-ordered, La2-xSrxNiO4, Excitations}

\maketitle

\section{Introduction}

Recent understanding of how charge-striped ordered phases containing significant disorder can produce hourglass-shaped magnetic excitation spectrum add further support to the charge stripe picture of hole doped cuprates\cite{boothroyd-Nature,Ulbrich}. With evidence for a normal phase charge-stripe order around 1/8 doping in yttrium based cuprates also highlighting the potential ubquitous nature of stripes beyond La based cuprates\cite{YBCOstripes}. These studies strongly motivate all efforts to improve our understanding of the charge-stripe ordered phase, to further our studies in  non-superconducting charge-stripe ordered materials such as La$_{2-x}$Sr$_{x}$NiO$_4$ (LSNO).

\begin{figure}[!ht]
\begin{center}
\includegraphics[width=8cm,clip=]{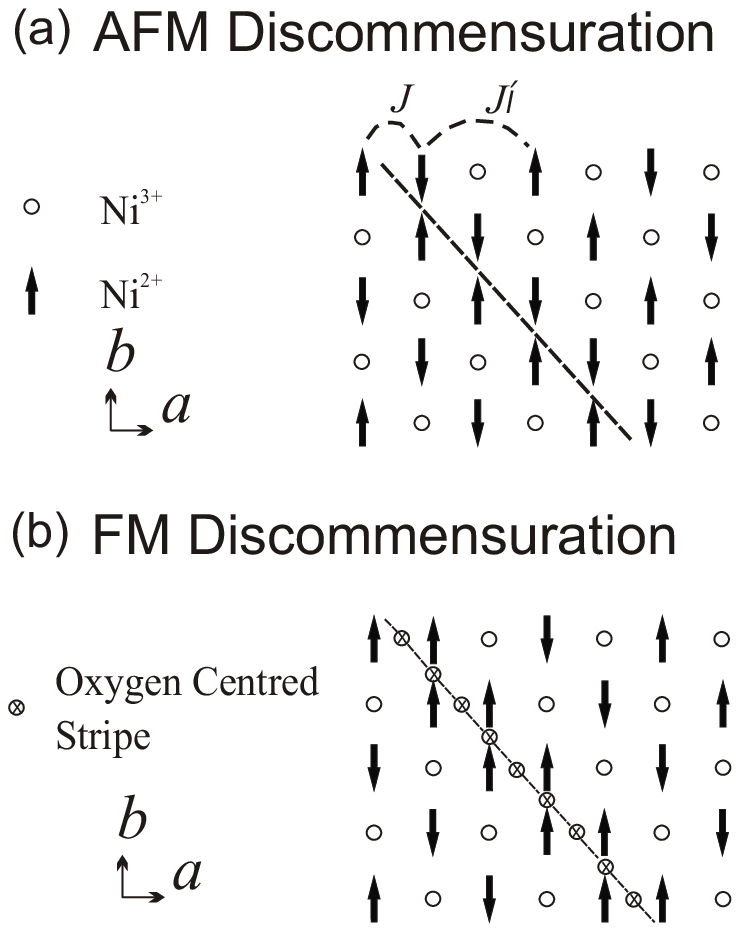}
\caption[Dispersion]{(Color online) Two types of discommensuration
that exist in La$_{2-x}$Sr$_{x}$NiO$_{4}$\cite{kajimoto-PRB-2003}.
The spins on nearest neighbor Ni sites have (a) antiparallel and
(b) parallel alignment, respectively, in the discommensuration
(indicated by a dashed line). Circles and arrows denote holes and
$S=1$ spins, respectively, on the Ni sites. The encircled crosses
in (b) denote the positions of an oxygen-centered charge stripe.
Discommensurations help stablize the magnetic order at half doping by increasing number of nearest neighbour exchange interactions $J$ between Ni$^{2+}$ ions, compared to the weaker next-nearest-neighbor interaction $J'$ across the Ni$^{3+}$ ions. In the $x = 0.5$ the two type of discommensurations are believed to be responsible for antiferromagnetic (a), and ferromagnetic(b) gapped excitation modes } \label{fig:discomm}
\end{center}
\end{figure}

The magnetic excitation spectrum of charge-stripe ordered LSNO, has been studied in detail for several doping levels, $ x = 0.275, 0.31, 1/3$ and $x = 0.5 $\cite{boothroyd-PRB-2003,bourges-PRL-2003,Woo,freeman-PRB-2005}.  Despite the success of modelling the excitation spectrums, significant differences in the role of discommensurations in the magnetic excitations of LSNO $x \sim 1/3$ and $x = 0.5$ are observed\cite{Woo,freeman-PRB-2005}. Discommensurations are variations in the charge stripe spacing, caused by pinning of the charge stripes to either the Ni or O sites of the Ni-O planes of LSNO. For doping levels near $x = 1/3$ discommensurations cause damping of the spatial extent of magnetic excitations. While the $x = 0.5$ magnetic excitation spectrum can be understood as a checkerboard charge order state containing discommensurations that lead to a spin stripe order, where discommensurations cause the development of new gapped magnetic excitation modes.  Figure \ref{fig:discomm} shows the two types of discommensurations that can exist in LSNO, that cause (a) antiferromagnetic, and (b) ferromagnetic excitations in the $ x  = 0.5$. The ferromagnetic or antiferromagnetic character of these modes is implied from the wavevector centre that they disperse away from with increasing energy transfer, being either a ferromagnetic or antiferromagnetic zone centres. We do however note that an alternative theory has been proposed to describe the antiferromagnetic excitation as due to stripe twining\cite{Yao-PRB-2007}. In this study we wish to address how discommensurations effect the magnetic excitation spectrum on doping towards $x = 0.5$, and determine the role, if any, of checkerboard charge order.

\section{Experiment Details and Results}

Single crystals of La$_{1.55}$Sr$_{0.45}$NiO$_4$ and  La$_{1.6}$Sr$_{0.4}$NiO$_4$ were grown using the floating-zone technique\cite{prab}. The  La$_{1.55}$Sr$_{0.45}$NiO$_4$ crystal was a rod of 6\,mm in diameter and 25\, mm in length  weighing  2.6\,g, and the La$_{1.6}$Sr$_{0.4}$NiO$_4$  crystal was a slab of dimensions $\approx 15 \times 10 \times 4$\,mm and weighed 1.8\,g . The samples used here are the same samples that were studied in our previous neutron diffraction measurements, which are reported elsewhere\cite{paul2,giblin-PRB-2008}.  Oxygen content of the La$_{1.6}$Sr$_{0.4}$NiO$_4$  was determined to be stoichiometric by thermogravimetric analysis,\cite{prab} whereas the results of the neutron diffraction study of La$_{1.55}$Sr$_{0.45}$NiO$_4$ are consistent with stoichiometric oxygen content\cite{giblin-PRB-2008}. The bulk magnetization of the $x = 0.45$ is consistent with that of an slightly oxygen deficient $x = 0.5$\cite{freeman-JSNM-2011},  but studying the $x = 0.45$ has the advantage of being well away from the half doped checkerboard charge ordered phase.

Neutron scattering were performed on the triple-axis spectrometers PUMA at FRM  II, and IN8 at the Institut Laue-Langevin. The data was collected with a fixed final neutron wavevector of $k_f = 2.662$\,\AA. A pyrolytic graphite (PG) filter was placed after the sample to suppress higher-order harmonic scattering. We measured the excitation spectrum of  the La$_{1.55}$Sr$_{0.45}$NiO$_4$ on PUMA, and the excitation spectrum of La$_{1.6}$Sr$_{0.4}$NiO$_4$ was measured on IN8. On both instruments the neutrons final and initial energy was selected by Bragg reflection off a double focusing pyrolytic graphite (PG) monochromator and analyzer respectively.  On PUMA the La$_{1.55}$Sr$_{0.45}$NiO$_4$ sample was mounted inside a cold cycle refrigerator, and on IN8 the La$_{1.6}$Sr$_{0.4}$NiO$_4$ sample was mounted inside an orange cryostat. Both samples were orientated so that  ($h$,\ $k$,\ 0) positions in reciprocal space could be accessed. In this work we refer to the tetragonal unit cell of LSNO, with unit cell parameters $a \approx 3.8$\ \AA , $c \approx 12.7$ \AA.

\begin{figure}\includegraphics[width=9cm]{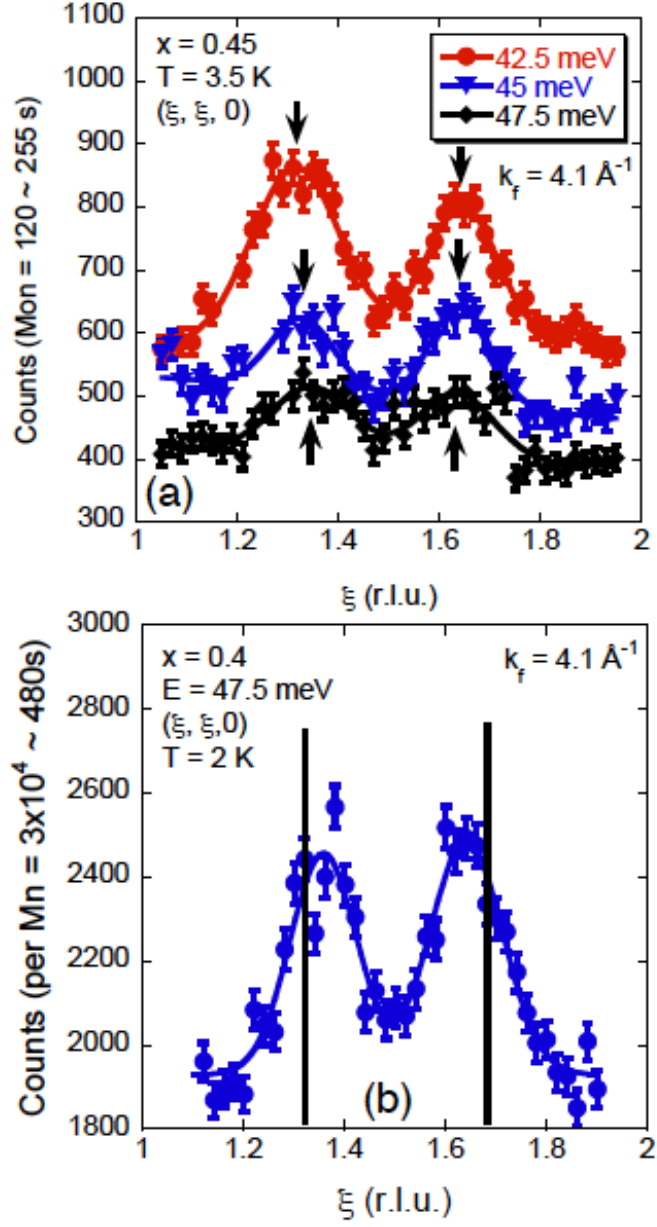} .\caption{ (color online) Constant energy scans  of the excitations from  La$_{2-x}$Sr$_{x}$NiO$_4$ $x = 0.45$ for $E > 40$\,meV measured on PUMA, and for b) $x = 0.4$ at 47.5 \,meV measured on IN8. Solid lines are used to indicate the result of a fit of two Gaussian lineshapes on a sloping background to the data. In (a) we indicated the centre of the Gaussian peaks by arrows to highlight the dispersion of this mode, while in (b) solid vertical lines are used to indicate the magnetic zone centre of the $x = 0.4$ } \label{first}
\end{figure}

In the $ x = 0.5$ we observed for $E >40$\,meV that the magnetic excitations of the acoustic magnon branch dispersed towards  ${\bf Q_ {AFM}} = (0.5,\ 0.5,\ 0)$ without a counter propagating mode\cite{freeman-PRB-2005}.  In figure \ref{first} we show the magnetic excitations from the ordered moments of LSNO $x = 0.4$ and $x = 0.45$ for $E >40$\,meV. The magnetic zone centres for the $x = 0.45$ and $x = 0.4$ are $(h + (1\pm \varepsilon) /2, k + (1 \pm \varepsilon) /2, 0)$, with $\varepsilon = 0.425$ and  $\varepsilon = 0.371$ respectively. For the $x = 0.45$ we see in Fig. \ref{first} (a) excitations that disperse towards $\bf {Q_ {AFM}}$ on increasing $E$ from 42.5\,meV to 47.5\,meV with no observable counter propogating mode, consistent with magnetic excitations of the $x = 0.5$. Similarly in Fig. \ref{first}(b) we observe magnetic excitations in the $x = 0.4$ are shifted towards the  $\bf {Q_ {AFM}}$ from the magnetic zone centre at $47.5$\,meV. It appears the magnetic excitations in LSNO lose their symmetry in the dynamic structure factor at a doping level between $x = 1/3$ and  $x = 0.4$, with this effect decoupled from checkerboard charge-ordering.

\begin{figure}\includegraphics[width=10cm]{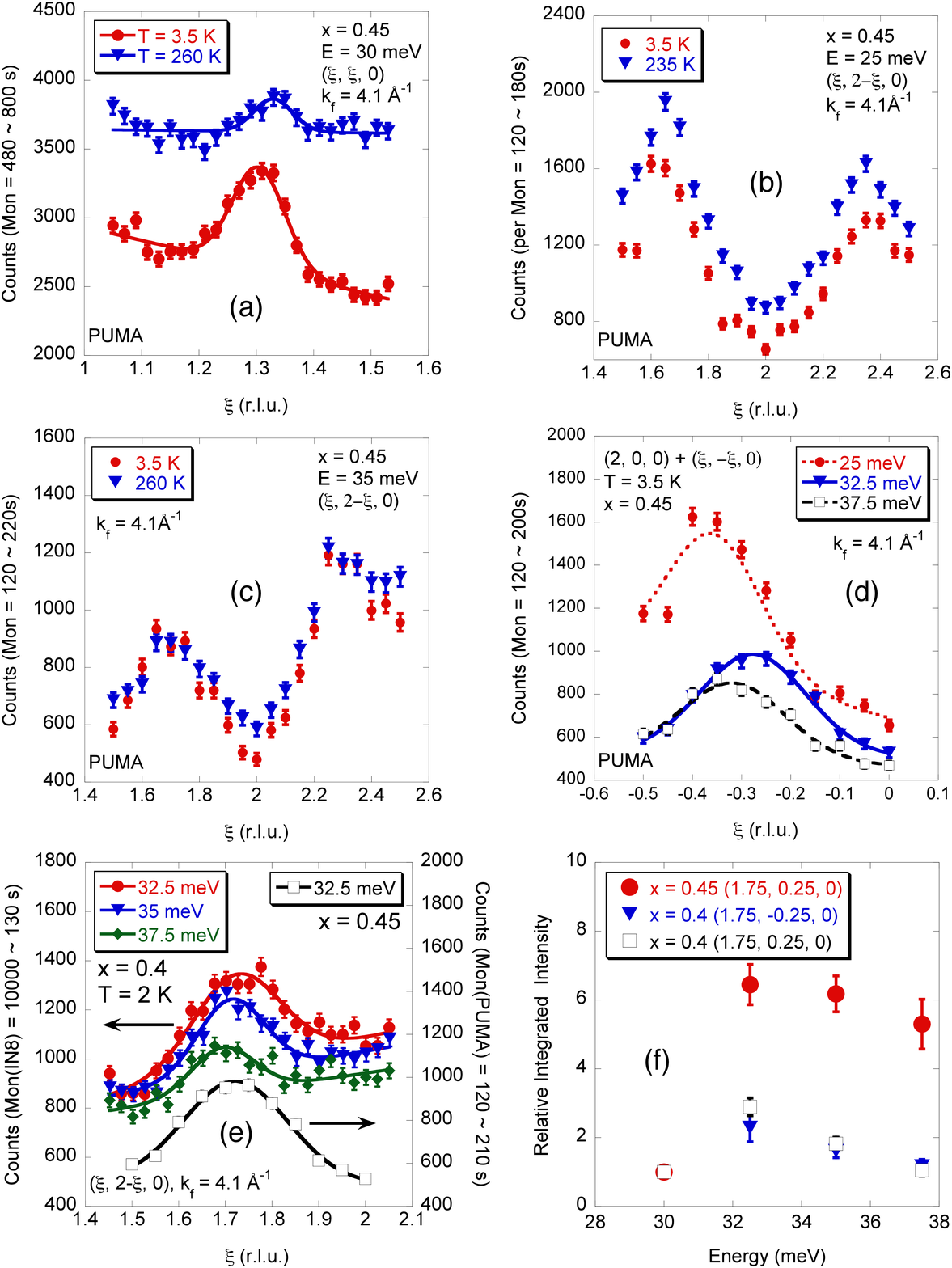} .\caption{ (color online) Constant energy scans  of the excitations from $x = 0.45$ (a) 30\,meV, (b) $E = 25$\,meV and (c) $E = 35$\,meV excitations measured at 3\,K and 260\,K. In (a) we show a scan through the acoustic magnon excitations, while (b) and (c) are scans through additional excitation modes. (d) Constant energy scans of  half a Brillouin zone at 25 meV, 32.5\,meV and 37.5\,meV of the $x = 0.45$ at 3\,K. 
In (a) and (d) the solid and dashed lines are results of a fit of a Gaussian peak on a sloping background to the data. There is no offset to the data shown in (a) - (c), and in (d) the 260\,K data was offset by the addition of 400 counts. (e) Constant energy scans of the excitations from $x = 0.4$  $E = 32.5$\,meV, $ 35$\,meV, and 37.5\,meV, with the equivalent constant energy scan of thw $x = 0.45$ at 32.5\,meV from (d) included for comparison purposes.  Solid, dashed and dotted lines in (a), (d) and (e) are the results of fits of  the data with a Gaussian lineshape on a sloping background. (f) A comparison of the intensity of the 32.5\,meV to 37.5\,meV excitations of the $x  = 0.4$  and $x = 0.45$, relative to the intensity of the acoustic magnon mode at 30\,meV measured at the wavevector indicated in (a).  The relative intensity at 30\,meV is one by definition, with the $x = 0.45$ data point at 30\,meV being obscured by that of the $x = 0.4$.} \label{ferromodes}
\end{figure}

Having studied the magnetic excitations from the acoustic magnon branch we searched for evidence of additional magnetic excitation modes. In figure \ref{ferromodes}(a)-(d)  we show constant energy scans over the energy range 25\,meV to 37.5\,meV in the $x = 0.45$. In this energy range in  La$_{2-x}$Sr$_{x}$NiO$_4$ magnon, phonon mixing in measurements by inelastic neutron scattering is a problem.
We first measured the excitations from the acoustic magnon dispersion at 30\,meV at 3.5\,K and 260\,K, as we show in Fig.\ref{ferromodes}(a). At this specific energy and wavevector we expect to observe no phonons in a constant energy scan\cite{boothroyd-PRB-2003,Woo,freeman-PRB-2005}. Consistent with a magnetic excitation measured above and below the magnetic ordering temperature, we observe a reduction in intensity of the acoustic magnetic excitation on increasing temperature from 3.5\,K to 260\,K due to the Bose effect.  In Fig.\ref{ferromodes}(b) we show a constant energy scan at 25\,meV parallel to $(1,\ -1,\ 0 )$ that passes through $(2,\ 0,\ 0)$ at 3.5\,K and 235\,K.  At 25\,meV the background increases with increasing temperature and the observed excitations appear to be temperature independent.  This scan passes over the wavevector of the acoustic magnon mode, at 235\,K the acoustic magnon will lose the same intensity as  in Fig .\ref{ferromodes} (a) in the scan of  Fig.\ref{ferromodes}(b).
Therefore to account for the apparent temperature independence of the scan in Fig.\ref{ferromodes}(b)
an additional excitation must exist at the same wavevector.  This excitation gains in intensity with
increasing temperature; this temperature dependence implies it should be assigned as a phonon excitation. In
Fig.\ref{ferromodes}(c) we show the same scan as in Fig.\ref{ferromodes} (b) but at 35\,meV . We observe the excitation to
lose intensity with increasing temperature, with the loss in intensity being larger than that
expected for the acoustic magnon branch. The additional excitation observed in Fig.\ref{ferromodes}(c)
loses intensity with increasing temperature implying a magnetic origin, consistent with the
additional ferromagnetic mode observed in the $x = 0.5$.

In figure \ref{ferromodes}(d) we show how we measured the dispersion of the additional magnetic excitation in the $x = 0.45$ with increasing energy transfer, and observe a dispersion apparently consistent with that observed in the $x = 0.5$\cite{freeman-PRB-2005}. After finding evidence of an additional magnetic excitation in the $x = 0.45$ sample, we searched for the same excitation in the $ x = 0.4$. We performed the same scans in reciprocal space in the $x = 0.4$ as in the $x = 0.45$ in Fig.\ref{ferromodes}(e). In  Fig.\ref{ferromodes}(e) there is an excitation in the $x = 0.4$ that appears to disperse, but there is a problem with this observation. To compare the excitations observed between  32.5\,meV to 37.5\,meV in the $x = 0.4$ and $x = 0.45$ we compare in Fig.\ref{ferromodes}(f) the fitted intensities obtained from fits of a Gaussian lineshape on a sloping background, relative to the intensity of the acoustic magnon at 30\,meV, without correcting for magnetic form factor or doping variation of the acoustic magnon at 30\,meV. In the $ x  = 0.45$ the intensity of the mode at $32.5$\,meV is a factor of 6.4$\pm$0.5 times greater than the acoustic magnon at 30\,meV, and at 37.5\,meV the excitation is still 5.3$\pm$0.7 times larger than the 30\,meV acoustic magnon. Whereas in the $x = 0.4$, the intensity of the mode at 32.5\,meV is a factor of 2.3$\pm$0.3 larger than the 30\,meV acoustic magnon, but by 37.5\,meV there is no intensity difference, with a  relative intensity of 1.2$\pm$0.2. Tentatively we conclude from the observed intensity variation, that there is a ferromagnatic excitation mode in the $x = 0.45$ but not in the $x = 0.4$, and that there is a presence of a flat phonon mode at 32.5-35\,meV at both doping levels. Within  this interpretation we can estimate the strengths of the different excitations present in the $x = 0.45$ at 32.5\,meV.  We know that between 30\,meV and 32.5\,meV the intensity of the acoustic magnon is small\cite{freeman-PRB-2005},  assuming this loss is negligble we use the measured intensity of teh acoustic magnon at 30 \,meV to estimate the relative intensities of the acoustic magnon and phonon measured at 32.5\,meV in the $x = 0.4$. Assuming the ratio between the acoustic magnon and phonon at 32.5\,meV has little doping variation, we estimate that   the ratio of acoustic magnon to optic phonon to ferromagnetic gapped excitation in the $x = 0.45$ at 32.5\,meV shown in Fig. 3 \ref{ferromodes}(d)   is 1:1.6:3.9.  We can, however, definitely say that a ferromagnetic
excitation rapidly gains intensity on doping from $x = 0.4$ to $x = 0.45$.

A second additional magnetic excitation was observed in the $ x = 0.5$ above the top of the acoustic magnon branch, and due to the mode centring it was interpreted as due to antiferomagnetic discommensurations\cite{freeman-PRB-2005}.  We searched inconclusively for this mode, due to the difficulty in establishing the maximum energy of the acoustic magnon dispersion, and not knowing the effect of charge ordering on the phonon modes at high doping levels\cite{Tranquada-PRL-2002} .

\begin{figure}\includegraphics[width=8.4cm]{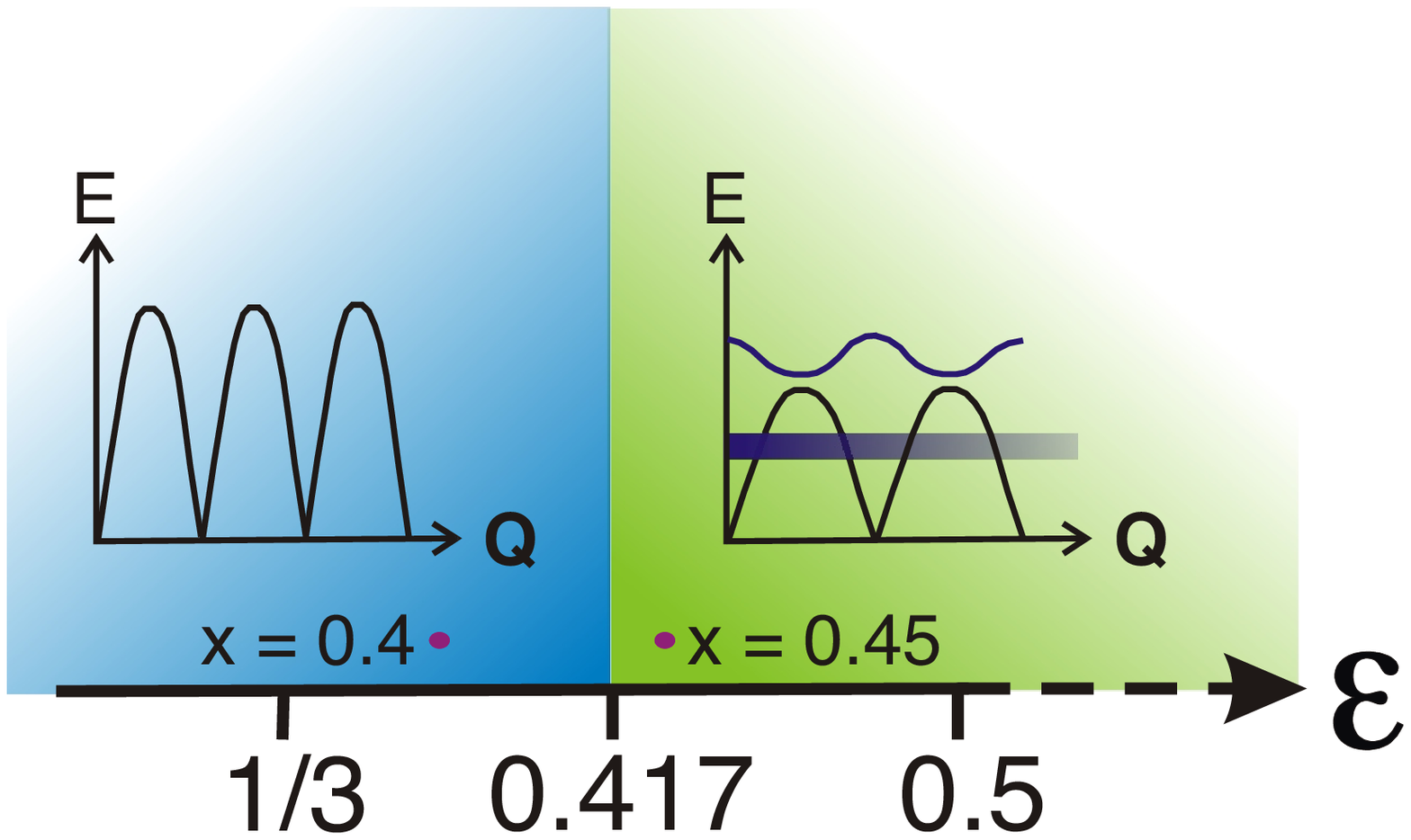} \caption{ A proposed phase diagram of the magnetic excitation spectrum from the ordered Ni$^{2+}$ S = 1 spins of LSNO for  $\varepsilon > 0.3$. Indicating in purple the additional antiferromagnetic and ferromagnetic magnetic excitation modes caused by discommensurations in the spin structure for for $\varepsilon > 0.417$.} \label{crossover}
\end{figure}

\section{CONCLUSIONS}

Our measurements of the magnetic excitation spectrums of the $x = 0.4$ and $x = 0.45$ appear to show two effects, with neither effect requiring the presence of checkerboard charge order.  The first effect is that the excitation spectrum at high energies ($> 40$\,meV) for both the $x = 0.4$ and $x = 0.45$  are not centred symmetrically about the magnetic zone centre, but are shifted towards ${\bf Q_ {AFM}} = (0.5,\ 0.5,\ 0)$. This contrasts with the symmetric spin wave cones observed in the $ x = 1/3$ and $x = 0.275$, with counter propagating modes that disperse with equal intensity towards and away from ${\bf Q_ {AFM}}$ in reciprocal space. Between $x = 1/3$ and $x = 0.4$ the only change in the materials we are aware of is the change in nature of the free charge carriers from electron like to hole like, but it is unclear how this should effect the magnetic excitations\cite{katsufji-PRB-1999}.

In the $x = 0.45$ an additional excitation mode is observed, that appears not to be present in the $ x = 0.4$.  In the $x = 0.45$, over the energy range 32.5 to 37.5\,meV we have shown evidence of a ferromagnetic excitation, which is  consistent with one of  two additional excitation modes observed in the $x = 0.5$. In the $x = 0.5$ the additional magnetic modes are qualitatively explained as arising from  discommensurations in the magnetic structure. That discommensurations produce new modes in the excitation spectrum of the $x = 0.45$ and not $ x = 0.4$, we tentatively propose is due to whether the discommensurations are distortions of the checkerboard spin state or the charge-stripe phase with stripes 3 Ni-Ni distances appart. That is, is the charge stripe order closer to the checkerboard order state or a stripe state with charge stripes three Ni sites apart, i.e. is $\varepsilon$ (= 1/charge order periodicity) $> 0.417$ or is $\varepsilon > 0.417$. For $\varepsilon < 0.417$ discommensurations only broaden the excitations, but for $\varepsilon < 0.417$ discommensurations lead to new excitation modes, as we depict in figure \ref{crossover}. Whether this means there will be a doping crossover in the magnetic excitation between charge stripe order with $\varepsilon = 0.25$ which has an optic magnetic excitation, and  $\varepsilon = 1/3$ which only has an acoustic magnetic excitation branch, we cannot say. But oxygen doped La$_2$NiO$_ {4.11}$ with $\varepsilon = 0.273$ appears to have an optic excitation mode, whereas the $x = 0.275$ and $x = 0.31$ appear not to have an optic magnetic excitation mode. As the number of magnetic excitation modes is a distingushing feature of theories of the magnetic excitations of the cuprates\cite{Seibold-PRB-2006}, we need to resolve the issue of whether the charge stripe ordered nickelates have the magnetic dispersion relation of their nearest commensurate charge-stripe structure or the dispersion of admixtures of different commensurate charge stripe ordered phases.


\begin{acknowledgments}
We wish to acknowledge that crystal growth for this work was supported by
the Engineering and Physical Sciences Research Council of Great Britain. This research project has been supported by the European Commission under the 6th Framework Programme through the Key Action: Strengthening the European Research Area, Research Infrastructures. Contract n�: RII3-CT-2003-505925.

\end{acknowledgments}


\begin{references}


\bibitem{boothroyd-Nature}
A. T. Boothroyd, P. Babkevich, D.  Prabhakaran, and P. G. Freeman, 2011 Nature  {\bf 375}, 561 (2011).

\bibitem{Ulbrich}
H. Ulbrich, P. Steffens, D. Lamago, Y. Sidis, M. Braden, arXiv:1112.1799

\bibitem{YBCOstripes}
David LeBoeuf, Nicolas Doiron-Leyraud, Julien Levallois, R. Daou, J.-B. Bonnemaison, N. E. Hussey, L. Balicas, B. J. Ramshaw, Ruixing Liang, D. A. Bonn, W. N. Hardy, S. Adachi, Cyril Proust, and Louis Taillefer Nature {\bf 450}, 533 (2007); David LeBoeuf, Nicolas Doiron-Leyraud, B. Vignolle, Mike Sutherland, B. J. Ramshaw, J. Levallois, R. Daou,
Francis Lalibert\'{e}, Olivier Cyr-Choini{\`{e}}re, Johan Chang, Y. J. Jo, L. Balicas, Ruixing Liang, D. A. Bonn,
W. N. Hardy, Cyril Proust, and Louis Taillefer, Phys Rev. B {\bf 83}, 054506 (2011); 
F. Laliberte, J. Chang, N. Doiron-Leyraud, E. Hassinger, R. Daou, M. Rondeau, B. J. Ramshaw, R. Liang, D. A. Bonn, W. N. Hardy, S. Pyon, T. Takayama, H. Takagi, I. Sheikin, L. Malone, C. Proust, K. Behnia, and L. Taillefer, Nature Communications {\bf 2}, 432 (2011); Tao Wu,
Hadrien Mayaffre, Steffen Kr�mer, Mladen Horvati{\'{c}}, Claude Berthier,
W. N. Hardy, Ruixing Liang, D. A. Bonn,and Marc-Henri Julien
Nature {\bf 477}, 191 (2011).




\bibitem{boothroyd-PRB-2003}
A. T. Boothroyd,  D. Prabhakaran, P. G. Freeman, S. J. S. Lister,
M. Enderle, A. Hiess, and J. Kulda, Phys. Rev. B {\bf 67},
100407(R) (2003).


\bibitem{bourges-PRL-2003}
P. Bourges, Y. Sidis, M. Braden, K. Nakajima, and J. M. Tranquada
Phys. Rev. Lett. {\bf 90} 147202 (2003).

\bibitem{Woo} H. Woo, A. T. Boothroyd, K. Nakajima, T.G. Perring,
C. D. Frost, P. G. Freeman, D. Prabhakaran, K. Yamada and J. M.
Tranquada, Phys. Rev. B {\bf 72}, 064437  (2005).


\bibitem{freeman-PRB-2005}
P. G. Freeman, A. T. Boothroyd, D.
Prabhakaran, C. D. Frost, M. Enderle, and A. Heiss, Phys. Rev. B
{\bf 71}, 174412 (2005).




\bibitem{Yao-PRB-2007}
D. X. Yao and E. W. Carlson, Phys. Rev. B \textbf{75} ,012414 (2007) .

\bibitem{kajimoto-PRB-2003}
R. Kajimoto, K. Ishizaka, H.
Yoshizawa, and Y. Tokura, Phys Rev B {\bf 67}, 014511 (2003).



\bibitem{prab} D. Prabhakaran,  Isla P., and A. T. Boothroyd, J.
Cryst. Growth \textbf{237}, 815 (2002)

\bibitem{paul2} P. G. Freeman, A. T. Boothroyd, D. Prabhakaran, M.
Enderle and C. Niedermayer, Phys. Rev. B \textbf{70}, 024413
(2004)


\bibitem{giblin-PRB-2008}
S. R. Giblin, P. G. Freeman, K. Hradil, D. Prabhakaran, and A. T. Boothroyd, Phys. Rev. B \textbf{78} ,184423 (2008) .

\bibitem{freeman-JSNM-2011}
P. G. Freeman, S. R. Giblin and D. Prabhakaran, J Supercond. Nov. Mag.  \textbf{24}, 1149 (2011).

\bibitem{Tranquada-PRL-2002}
J. M. Tranquada, K. Nakajima, M.  Braden, L. Pintschovius, and
R. J. McQueeney, Phys. Rev. Lett. {\bf  88}, 075505 (2002).



\bibitem{katsufji-PRB-1999}
T. Katsufuji, T. Tanabe, T. Ishikawa, S. Yamanouchi, Y. Tokura, T. Kakeshita, R. Kajimoto, and H. Yoshizawa, Phys. Rev. B {\bf 60}, R5097 (1999)


\bibitem{Seibold-PRB-2006}
G. Seibold,  and J. Lorenzana, Phys. Rev. B  {\bf 73}, 144515 (2006).














\end{references}
\end{document}